\documentclass[manuscript, nonacm]{acmart}

\usepackage{float}
\usepackage{xspace}
\usepackage{soul}
\usepackage[dvipsnames]{xcolor}
\DeclareRobustCommand{\systemName}{\textsc{Roop}\xspace}

\AtBeginDocument{%
  }

\setcopyright{none}
\renewcommand\footnotetextcopyrightpermission[1]{}
\begin{document}

% \title{[TEMP TITLE] CHI 2027 ADOBERESEARCH}

\title{Who Shaped What? Untangling Human and AI Creative Contributions Through Spatial Provenance in Image Editing}
% Untangling Human and AI Creative Contributions Through Spatial Provenance in Image Editing
% What Provenance Captures About Creatie Contribution in Human-AI Image Editing
% Interpreting Human and AI Creative Contributions Through Spatial Provenance in Image Editing

% Tracing Human and AI Creative Contributions in AI-Edited Images Through Spatial Interaction Provenance
% Pentimento: Tracing Human and AI Creative Contributions in Images Through Spatial Interaction Provenanc
% Pentimento: Tracing Human and AI Creative Contributions Through Spatial Interaction Provenance
%Thought Provenance: Tracing Creative Contribution in Human-AI Image Editing
% Thought Provenance: Measuring Creative Ownership in Human-AI Image Editing
% Thought Provenance: Capturing Creative Initiative in Human-AI Visual Editing
% Thought Provenance: Capturing Creative Initiative in Human-AI Visual Editing
% Tracing Creative Initiative in Human-AI Visual Editing
% \systemName: A Proposal-First Interface for Capturing Thought Provenance in Human-AI Visual Editing

% Ideas Leave Traces: Thought Provenance for Human-AI Creative Collaboration

\author{Riya Sahni}
\authornote{This work was done while the author was an intern at Adobe Research.}
\affiliation{%
  \institution{Columbia University}
  \city{New York City}
  \country{USA}
}
\email{riya.sahni@cs.columbia.edu}

\author{Christine Dierk}
\affiliation{%
  \institution{Adobe Research}
  \city{San Jose}
  \country{USA}
}
\email{dierk@adobe.com}

\author{Lydia Chilton}
\affiliation{%
  \institution{Columbia University}
  \city{New York City}
  \country{USA}
}
\email{chilton@cs.columbia.edu}

\author{Eunyee Koh}
\affiliation{%
  \institution{Adobe Research}
  \city{San Jose}
  \country{USA}
}
\email{eunyee@adobe.com}

\author{Doga Dogan}
\authornote{These authors jointly supervised this work.}
\affiliation{%
  \institution{Adobe Research}
  \city{San Jose}
  \country{USA}
}
\email{doga@adobe.com}

\author{Tongyu Zhou}
\authornotemark[2]
\affiliation{%
  \institution{Adobe Research}
  \city{San Jose}
  \country{USA}
}
\email{tongyuz@adobe.com}

\renewcommand{\shortauthors}{Sahni et al.}

\begin{abstract}

Images co-edited with AI are often labeled to disclose AI use. However, creators feel these labels don’t sufficiently describe their contributions to the editing process. Editing histories preserve requests, suggestions, and revisions, but how do creators use these logs to judge contribution? We investigated this with \systemName, a mixed-initiative design probe that links image regions to editing histories (spatial provenance) and summarizes interactions in a heat map. Twenty participants edited images, annotated human and AI contributions, then reviewed the spatial provenance heat map. Participants credited themselves for directing changes while crediting AI for determining visual details. They judged contribution partly by whether their interactions achieved their intentions. Reviewing provenance helped participants remember interactions without always changing whom they credited. These findings motivate provenance systems that preserve enough context for creators to explain how interactions shaped their work, alongside the record of what happened.

\end{abstract}

\begin{teaserfigure}
  \centering
  \includegraphics[width=\textwidth]{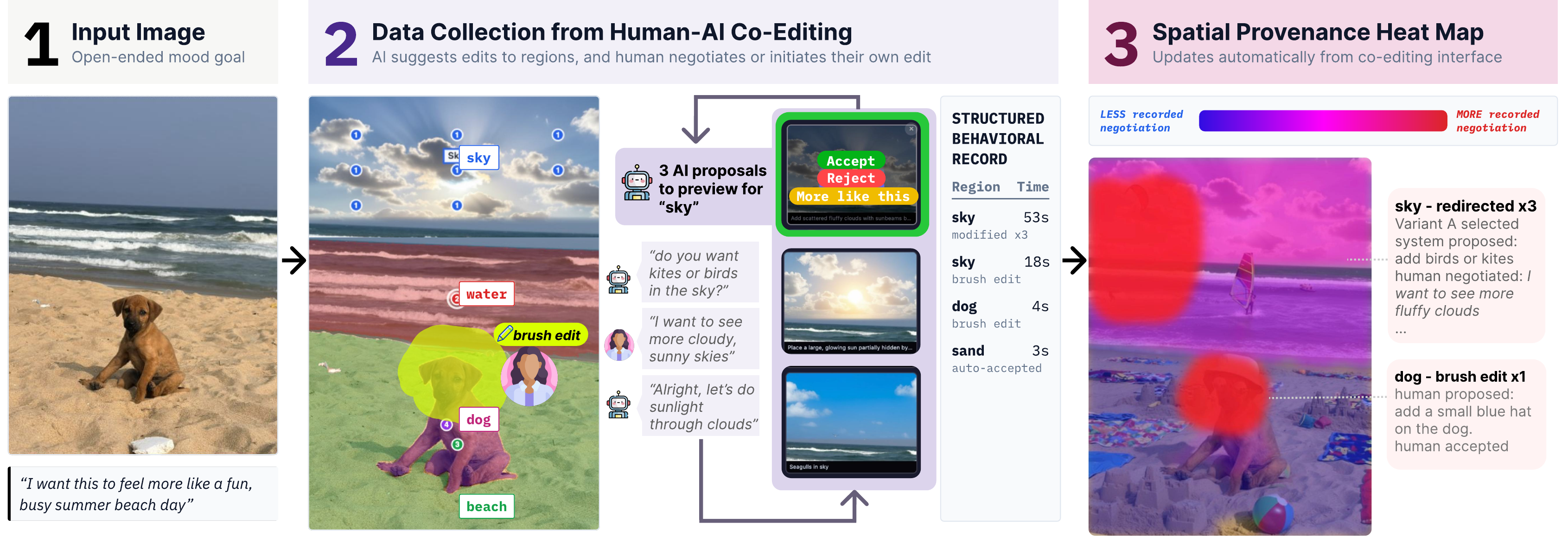}
  \caption{\systemName design probe. (1)~The user uploads an image and enters a mood goal. (2)~The user explores AI suggestions, asks for revisions, or makes their own requests using the brush tool. Accepted edits inform later suggestions, and \systemName records these interactions by region. (3)~The heat map shows which regions involved more or less recorded negotiation within the session. The user can hover over a region to see the suggestions, their responses, and the options they previewed. We use the heat map as a probe and not a definite measure of creative contribution.}
  \label{fig:teaser}
\end{teaserfigure}

\maketitle

% %% The acknowledgments section
% \begin{acks}
% \end{acks}
% \includegraphics[width=\textwidth]{figures/teaser}

\section{Introduction}
Generative AI can now participate as a capable collaborator throughout the development of creative work, including image editing and writing \cite{shen_empowering_2024, lee_coauthor_2022}. It can propose edits, suggest directions, and influence what gets created before the creator has fully decided what they want \cite{fan_contextcam_2024, suh_luminate_2024}. For artists and other creators, being able to understand and explain their own contributions can affect how their work is credited, attributed, and understood \cite{he2025contributionsdeservecreditperceptions, hoque2024hallmark}. Creators may therefore need to show not only where AI was used, but which AI-generated changes came from their own ideas and which directions developed through collaboration with AI. This matters not only for how creators understand their own role, but also for how others interpret and evaluate AI-assisted work when the final image or a coarse AI-use label does not reveal how it was made.

An AI-generated edit can differ widely depending on how proactive the creator was during the process. A creator might already know what they want and ask AI to execute their instructions verbatim. In another case, AI might suggest a new direction that the creator then accepts, refines, or develops further \cite{rezwana_cofi_2023, lin_beyond_nodate}. The visible outcome may look similar in both cases, even though the creator and AI played different roles in deciding what should change. A record of the editing history can preserve the prompts, suggestions, revisions, and selections that led to the result \cite{lee_coauthor_2022, hoque2024hallmark, guo_prompthis_2024}, but these actions do not by themselves explain how the creative direction itself was developed.

Prior work has shown that people consider several factors when judging human and AI contribution, including what each party contributed, how much they contributed, and who took the initiative \cite{he2025contributionsdeservecreditperceptions}. However, it remains unclear how provenance systems could be designed to capture this nuance, and how low-level interaction logs map to a creator's perception of contribution.
% \textcolor{blue}{However, we know less about whether the interactions recorded during creation capture the same aspects of contribution that creators themselves use when making those judgments.}
% \hl{We build on this work by studying how creators explain their own and AI’s contributions to work they actually created together. }
We study this by comparing creators' explanations of their own and AI's contributions with a provenance record of the same creative process.

% Given the growing use of generative AI in image-editing workflows [CITE], we focus on AI-assisted image editing. 
One creative practice where the contribution line between human and AI is often blurred is AI-assisted image editing. In illustration or writing, where provenance has been more extensively studied~\cite{gjing2026man, he2025contributionsdeservecreditperceptions, hoque2024hallmark}, personal style typically persists across the whole piece and remains legible even with AI assistance. Image editing, however, often involves discrete and mechanical operations (inpainting, background removal, etc.) that carry no stylistic signature of their own, so there is no intrinsic marker of which parts were AI-generated. 
% giving  so contribution is usually legible in the work itself. Many image editing operations are different; applying a local edit or changing the sky color are convention-following changes that carry little personal signature because they look the same whether a person or AI executed it. 
This ambiguity is what makes image editing a compelling setting to study
% Prior work has shown that connecting editing history to the parts of an image that changed can help people inspect how those regions were produced or revised [CITE]. This makes image editing a useful setting 
for comparing creators’ own judgments and interpretations with the recorded history associated with the same parts of their image.

Specifically, we ask:
\begin{itemize}
    \item \textbf{RQ1:} How do creators understand their own and AI’s creative contributions during AI-assisted image editing, and how are those judgments affected when AI suggests what to edit?
    \item \textbf{RQ2:} What can a record of human–AI editing interactions capture about creators' judgments of their own contribution, and what does it miss?
\end{itemize}

To investigate these questions, we developed \systemName, a mixed-initiative image-editing design probe. \systemName supports both creator-initiated editing and a workflow in which AI can suggest editing directions that the creator can negotiate. It records the resulting interactions, attaches them to the regions of the image they affect, and summarizes that history spatially over the final image as a \textit{spatial provenance} heat map. Instead of situating Roop as a definitive measure of creative contribution, we use it as probe to compare creators' own understanding of contribution with the recorded interaction history around different parts of the image.

We conducted a within-subjects study with 20 participants comparing user-initiated and AI-suggested image editing, then examined how participants’ own judgments of their contributions compared with \systemName's spatial provenance record. By comparing creators’ annotations with the histories of their own edits, we examine how they connect recorded interactions to creative decisions. 14 of 20 participants rated their contribution lower when AI suggested what to edit. Participants distinguished contribution based on who introduced an idea, who shaped its direction, how much they deliberated, and what they selected or refined. Comparing these judgments with \systemName's record showed that provenance could recover interactions participants had forgotten, but recorded activity did not map cleanly onto creative contribution. Similar interaction histories could reflect correcting AI toward an existing idea or exploring a direction shaped through AI suggestions. Earlier AI suggestions could also influence later human-initiated edits, and a single edited region could contain decisions attributed partly by the creator and partly by AI.

Together, these findings distinguish identifying interactions from interpreting its creative significance. Connecting an interaction history to an image region helped participants revisit what happened, but did not fully explain which decisions they attributed to themselves or AI, or how those decisions informed their overall judgment. Some disagreements concerned the heat map’s scoring and display. In other cases, participants recognized their recorded actions but questioned what those actions should count toward their contribution. This motivates provenance systems that preserve not only what happened, but enough context for creators to interpret how different interactions shaped the work.

Our contributions are:

\begin{itemize}
    \item \textbf{Roop, a mixed-initiative design probe} that links human–AI editing interactions to image regions, enabling creators to compare their own contribution judgments with the recorded history of the same edits.
    \item \textbf{An empirical account} of how creators attributed human and AI contributions within their own edits, including how different creative decisions could coexist within the same edited region and develop across multiple interactions.
    \item \textbf{Design implications} that help creators reconstruct and communicate contribution without treating recorded activity as a direct measure of authorship or credit.
\end{itemize}

\section{Related Work}

We draw on three areas of prior work: (1) how people judge contribution in work created with AI, (2) how creative processes are recorded and presented through provenance, and (3) the distinction between recorded actions and the reasoning behind creative decisions.
\subsection{Judging Creative Contribution in Human-AI Creation}
When people create with generative AI, the human and AI can play different roles in shaping the work. Mixed-initiative frameworks describe settings where both parties can introduce changes or creative directions \cite{rezwana_cofi_2023, lin_beyond_nodate}. Creative systems have supported these roles in different ways: AI can propose ideas or alternatives that creators evaluate and refine \cite{fan_contextcam_2024, suh_luminate_2024, huh_videodiff_2025}, while creators can steer generated content, revise intermediate results, and select which outputs to incorporate \cite{lee_coauthor_2022, wu_aichains_2022}. Studies of longer-term creative practice also show that creators can shift how they use and respond to AI as their work develops \cite{tang_how_2026}.

In this paper, we use \textit{creative contribution} to refer to the ideas, decisions, and judgments through which a human or AI shapes what the work becomes. This focus builds on prior work examining different types of contribution and cognitive roles involved in human-AI creation \cite{he2025contributionsdeservecreditperceptions, wan_coco_2024}. We draw on work on ownership, authorship, agency, and attribution where it helps explain how people judge contribution, but we do not treat all these concepts as equivalent. For example, psychological ownership refers to the feeling that something is "mine'' \cite{pierce_psychological_2003}, while authorship and attribution concern who is recognized or credited for producing a work. Recent HCI work has similarly argued for distinguishing different meanings of creative ownership rather than treating ownership as a single construct \cite{polimetla_paradigm_2026}.

Prior work shows that people consider several parts of the creative process when making these judgments about creative contribution. He et al. presented participants with scenarios in which human and AI collaborators differed in contribution type, contribution amount, and whether a contribution was initiated proactively by the AI or in response to a request \cite{he2025contributionsdeservecreditperceptions}. Draxler et al. found that greater user influence over AI-generated text increased participants' sense of ownership, even though ownership and declared authorship did not always align \cite{draxler_ghostwriter_2024}. Xu et al. similarly found that people reasoned about ownership of human-AI creations through factors including originality, effort, and control \cite{xu_mine_2024}. Other studies connect experiences of agency and ownership to the control that creators retain over AI suggestions and the ways they refine, appropriate or respond to generated content \cite{rafner_agency_2025, carrera_where_2026}.

Other work has examined what happens when contribution is represented more explicitly. ArtSplit asked artists to respond to a provotype that quantified human and AI activity across stages of a creative workflow. Artists challenged the idea that measurable activity could directly represent creative intent, agency, or ownership \cite{gjing2026man}. CoCo Matrix instead looks at the different cognitive contributions that occur during human-AI writing, beyond the amount of activity or the final output alone \cite{wan_coco_2024}.

Together, this work shows that people do not judge creative contributions from the amount of human or AI activity alone. Prior work has studied these judgments through predefined contribution scenarios \cite{he2025contributionsdeservecreditperceptions}, retrospective accounts of human-AI creations \cite{xu_mine_2024}, think-aloud and interview studies of AI-assisted creation \cite{rafner_agency_2025}, and representations that explicitly categorize or quantify contribution \cite{gjing2026man, wan_coco_2024}. 

Our study brings these questions into a comparison between creators’ explanations and the recorded histories of \textit{edits they made themselves}. Since participants had made the edits, they could explain not only what actions occurred, but what they had been trying to achieve and how they understood those actions as contributing to the result. He et al. examined attribution through hypothetical scenarios, while ArtSplit elicited responses to a depicted artist’s process \cite{he2025contributionsdeservecreditperceptions,gjing2026man}. Other studies have examined ownership during actual AI-assisted creation, including writing with provenance support \cite{draxler_ghostwriter_2024,hoque2024hallmark}. We ask participants to annotate their contributions before reviewing their own regional editing histories, then examine what they remember and how they interpret those interactions. This allows us to distinguish reconsidering what happened from reconsidering how that participation should count toward contribution.

\subsection{Provenance and Representation of Human-AI Creative Processes}

Provenance systems record information about how digital content was created and changed. Standards like C2PA can preserve actions performed on an asset, including its creation and subsequent edits, while maintaining a verifiable history of those changes \cite{c2pa_spec}. Prior work has also proposed preserving generative histories, alternatives, and creative choices as evidence of how creators exercised control when producing work with AI \cite{huang_copyrighting_2025}. Huang et al. examine this question through copyright law, using the distinction between an idea and its expression to reason about whether expressive elements were determined by the human or supplied by AI \cite{huang_copyrighting_2025}. Our study instead examines how creators themselves make these distinctions when judging contribution to edits they produced, and how those judgments compare with a regional provenance record of the same process.

HCI has explored several ways of preserving and presenting creative process histories. HaLLMark records and visualizes interactions between writers and LLMs to support provenance and transparency \cite{hoque2024hallmark}. PrompTHis preserves relationships between prompts and generated images so users can review how their prompting process developed and how changes to prompts affected the resulting images \cite{guo_prompthis_2024}. DesignTrace Captures the progression of designers' visual and semantic ideas through command histories, state tracking, and branching structures, allowing creators to revisit earlier decisions and relationship among alternatives \cite{peng_designtrace_2026}. Other work has similarly used intermediate artifacts and version histories to support reflection on how creative work develops over time \cite{rawn_quickpose_2023, sterman_kaleidoscope_2023, hammad_tracing_2026}.

Image-generation and editing systems have also connected editing history to the visual content being changed. HistoryPalette organizes previous image alternatives by spatial position, topic, and creation time so creators can revisit and reuse earlier work \cite{benharrak_historypalette_2026}. Inverse Image Editing represents editing history through region-level operations that support re-editing and revision control \cite{hu_inverse_2013}. These systems show that editing history can be tied to particular parts of an image rather than presented only as a chronological sequence.

Our probe builds on this idea by attaching recorded human-AI interactions to the regions of an image they affected. We use this spatial provenance record as a concrete representation that creators can inspect and compare with their own understanding of contribution.

\subsection{From Recorded Interactions to Creative Contribution}

Research on \textit{design rationale} has long examined how to preserve not only the changes made during a creative process, but also the reasons, alternatives, goals, and criteria  behind those changes \cite{shum_rationale_1994, mckerlie_rationale_1994, carroll_rationale_2003}. More recent HCI work shows a similar limitation with contemporary creative tools: interaction and command histories can show what a designer did while leaving important goals and rationales unavailable without additional explanation \cite{cheng_lost_2026}. Work on creative activity traces similarly examines the different kinds of process information that traces can preserve and how those traces can be represented \cite{hammad_tracing_2026}.

This limitation also matters when provenance is used to describe human and AI creative contribution. A record can show that AI suggested an edit, the creator revised it, and the creator selected the final result. Those interactions tell us how the edit developed, but they do not by themselves determine how much proposing the idea, revising it, or choosing the result mattered to the creator's understanding of contribution.

Together, these areas of work establish that contribution judgments involve more than identifying recorded actions. What remains useful to examine is how creators explain the relationship between the two when reviewing their own work. In mixed-initiative image editing, we compare participants’ initial annotations with their explanations during provenance review. We examine both discrepancies concerning the representation and cases where participants recognize their actions but interpret their contribution differently from what the display suggests.

\section{\systemName: Probe Design}
\label{sec:probe}
% \textcolor{red}{[need to justify the proposal-first editing interface as the best type of system to capture provenance data and inform the provenance heat map. why do we need proposal-first editing? why not just attach the provenance log or like collect data from existing human-ai image editing workflows/tools? the answer is because provenance in those cases is very one-sided: it's always the human initiating a creative edit. on the flip side, creative ownership \& provenance becomes more complicated when creative ideas are being sparked and molded from both side - the human and the AI. current image editing tools are headed in this direction, and yet we don't fully understand how humans perceive ownership with this new interaction type. we developed \systemName, a proposal-first image editing tool that allows both AI and the human to lead creative directions, so we could measure and understand perceptions of creative ownership as AI becomes a more active creative collaborator. ]}

We developed \systemName to examine how creators understand their contributions when both they and AI can introduce editing directions, and how those judgments relate to a record of the editing process. The probe combines AI-suggested editing, user-initiated regional edits, and a spatial interaction history that creators can inspect alongside their image. Together, these capabilities let participants make creative decisions and later discuss those decisions in relation to the interactions recorded around the same regions.

To inform \systemName's design, we conducted formative interviews with 6 professional creatives who regularly used AI in their image-editing workflows: 2 professional artists and 4 professional designers. They described wanting to communicate \textit{how} they worked with AI, beyond indicating just \textit{whether} AI was used. These informed two design goals:
% \textcolor{blue}{background/demographic / any main themes identified?}
% who use AI editing tools. All 6 creators felt that existing disclosure tools failed to capture the nuance of their creative process. They wanted to show \textit{how} they worked with the AI and not just \textit{whether} they did, which is a shared sentiment across the broader creator community [META 2024 ARTICLE]. This pointed to two design goals:
\begin{enumerate}
    \item \textbf{Record how editing directions develop.} Preserve suggested edits, creators' responses, and subsequent revisions as part of the editing workflow.
    \item  \textbf{Connect the record to the image.} Allow creators to inspect the history associated with a particular region and discuss how it relates to their contribution.
\end{enumerate}

\systemName links its records to image regions so that the edited image and the interactions associated with an edit can be examined together. It also provides a heat map summarizing recorded engagement across regions (Figure~\ref{fig:architecture}). We use this representation as a probe for eliciting creators' interpretations of contribution, rather than as a validated allocation of authorship or credit. We refer to the linked record as spatial provenance: it preserves expressed suggestions and responses through which creative decisions develop. We used a heat map, because color provides an overview of relative negotiation intensity, and hovering reveals the history behind each region’s score. This lets creators move between the spatial summary and the specific interactions when comparing the record with their own annotations.

% \begin{figure*}[t]
%     \centering
%     \includegraphics[width=\textwidth]{figures/v1-technical-diagram.pdf}
%     \caption{\systemName probe structure.
%     The editing workflow supports both AI-suggested and user-initiated directions so participants can negotiate how edits develop. The resulting suggestions, revisions, previews, and decisions are recorded by region and later presented through the spatial provenance heat map for reflection. Roop is used as a design probe rather than evaluated as a technical system.}
%     \Description{A two-part system diagram. The upper panel shows
%     image and mood input, planning with GPT-4o, segmentation with
%     SAM3, regional suggestions, preview generation with Nano Banana,
%     creator review, and image updates. Feedback loops support
%     requesting alternatives and generating subsequent suggestions.
%     Brush edits provide an alternative input, and optional
%     harmonization produces a finished image. A logging connection
%     leads to the lower panel, where saved regional interactions
%     are scored and ranked to produce a heat map with editing
%     histories available on hover.}
%     \label{fig:architecture}
% \end{figure*}

\begin{figure}[!ht]
\vspace{-6pt}
  \centering
  \includegraphics[width=0.8\textwidth]{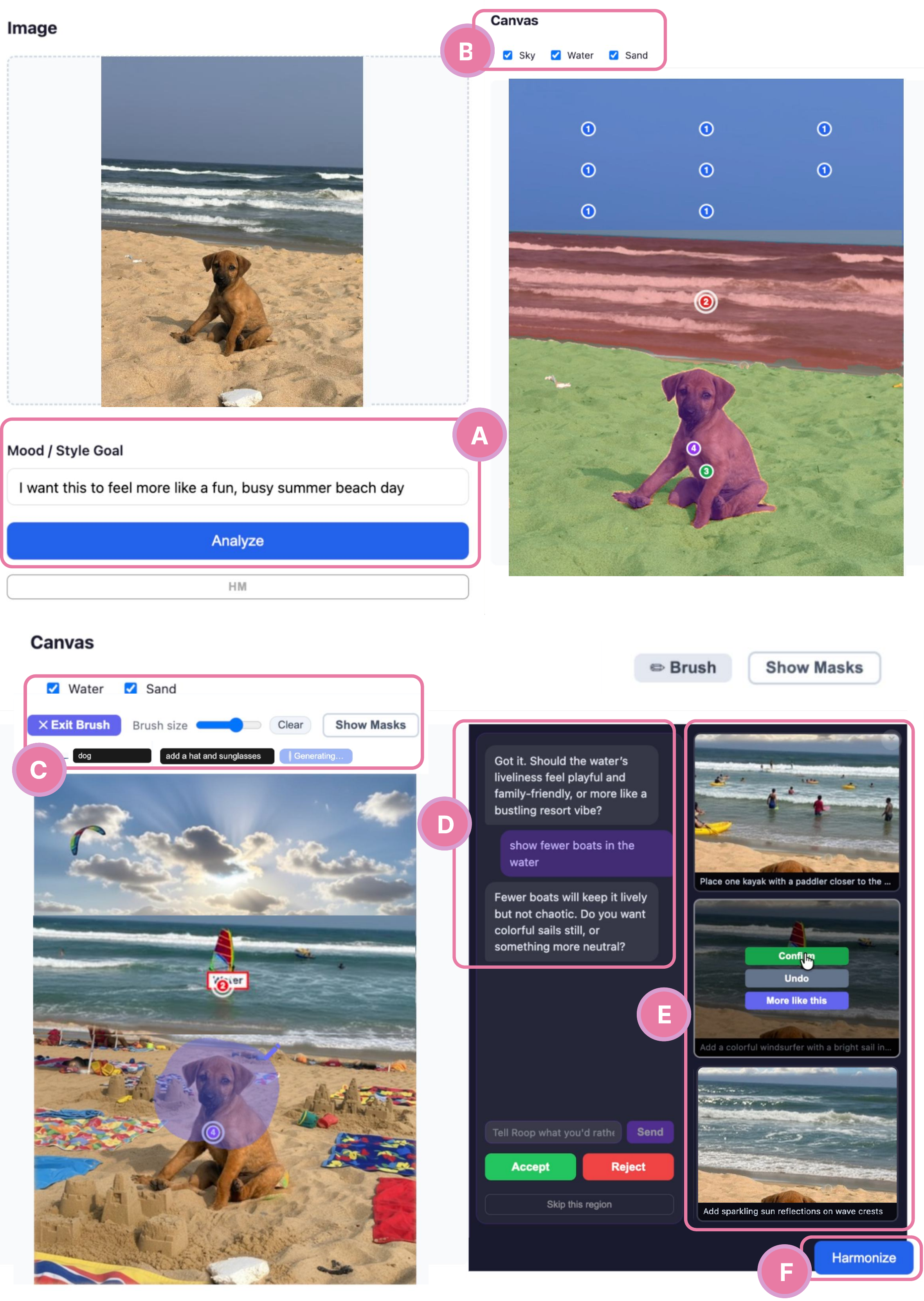}
  \caption{\systemName probe walkthrough. (A)~The user uploads an image and enters an open-ended mood goal; clicking Analyze triggers segmentation and proposal generation. (B)~\systemName segments the image into negotiable regions and overlays AI-generated edit proposals, each labeled by region. (E)~The user previews generated alternatives for the selected region on the canvas. (D)~The conversational negotiation panel lets the user accept, reject, redirect via chat, or request more variants. (C)~The brush tool lets the user paint a region and specify their own generative edit without first choosing an AI proposal. (F)~The Harmonize button generates options for blending the regional edits into a coherent image.}
  \label{fig:system1}
  \vspace{-6pt}
\end{figure}

\begin{figure}[!ht]
\vspace{-6pt}
  \centering
  \includegraphics[width=0.8\textwidth]{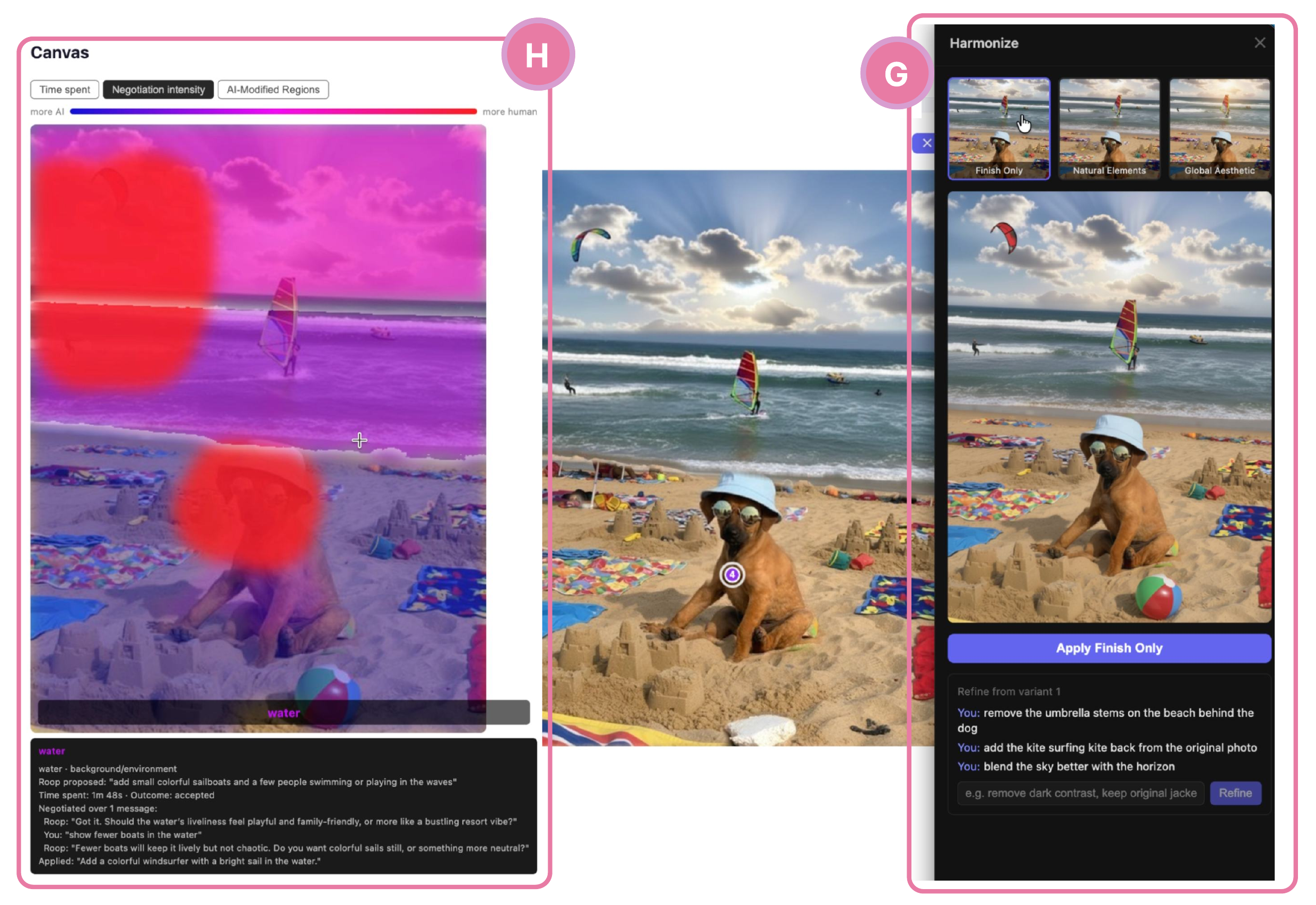}
  \caption{\systemName probe walkthrough (continued). (G)~The Harmonize panel presents three blending options at different levels of aggressiveness; the user selects a result and can refine further via chat. (H)~The spatial provenance heat map overlays the output image; warmer colors indicate regions higher ranked in negotiation intensity within the session, while cooler colors indicate lower ranked intensity. Brush-initiated edits receive maximum intensity. Hovering over a region reveals its recorded suggestions, user responses, and previews, allowing the creator to inspect how the edit developed.}
  \label{fig:system2}
  \vspace{-6pt}
\end{figure}
We illustrate the workflow through Sophia, a content creator editing a beach photo for a summer campaign. She wants the beach photo to feel lively and festive but has not decided which specific edits to make. She uploads the image, a mood goal, and clicks \textit{Analyze} ~(Figure~\ref{fig:system1} A).

% Most creators - especially novice creators - cannot clearly specify what they want, before any output exists to react to [cite our formative work]. Existing editing tools allow users to select specific regions and prompt changes for an AI to implement, placing the full cognitive burden of creative specification on the user upfront [CITE]. \systemName inverts this sequence: AI proposes edits based off its interpretation of the user's mood goal, and users can negotiate these individual proposed edits within the context of the whole image. This design enables the thought provenance record, because it captures every negotiation decision as a byproduct of the human's normal editing interaction with the AI. 

\subsection{Editing through AI suggestions and user-initiated changes}

\systemName supports both responding to AI suggestions and initiating an edit directly. This allows editing interactions to begin with either human or AI within the same editing session.

\subsubsection{AI-suggested regional edits}

After Sophia clicks Analyze (Figure \ref{fig:system1}~A), GPT-4o uses the image and mood goal to generate a structured editing plan. The plan identifies regions that could be changed, groups them by visual layer, and describes possible edits. SAM3 localizes the proposed regions in the image (Figure \ref{fig:system1}~B). In Sophia’s example, these include the sky, water, sand, and dog.

When Sophia opens a region, GPT-4o generates two or three edit suggestions. Regional and brush-initiated edits were generated using Google's Gemini 2.5 Flash Image (Nano Banana), accessed through Adobe’s Firefly image API\footnote{\url{https://developer.adobe.com/firefly-services/docs/firefly-api/}}. The image model receives the full working image and an instruction describing the proposed regional change, with instructions to preserve the surrounding content.

Sophia can preview a suggestion on the canvas (Figure \ref{fig:system1}~E), apply it, and confirm or reject it. She can also request more similar options or redirect the suggestions through chat (Figure \ref{fig:system1}~D). For example, she might ask for fewer boats in the water and inspect the resulting alternatives before choosing one. \systemName records the suggestions, chat responses, previews, and decisions in the history associated with that region.

This workflow records more than the accepted output. It preserves the alternatives and responses through which Sophia reached that result, allowing her to revisit how the edit developed.

\subsubsection{Suggestions respond to earlier edits that were made} 

\begin{figure*}[t]
    \centering
    \includegraphics[width=\textwidth]{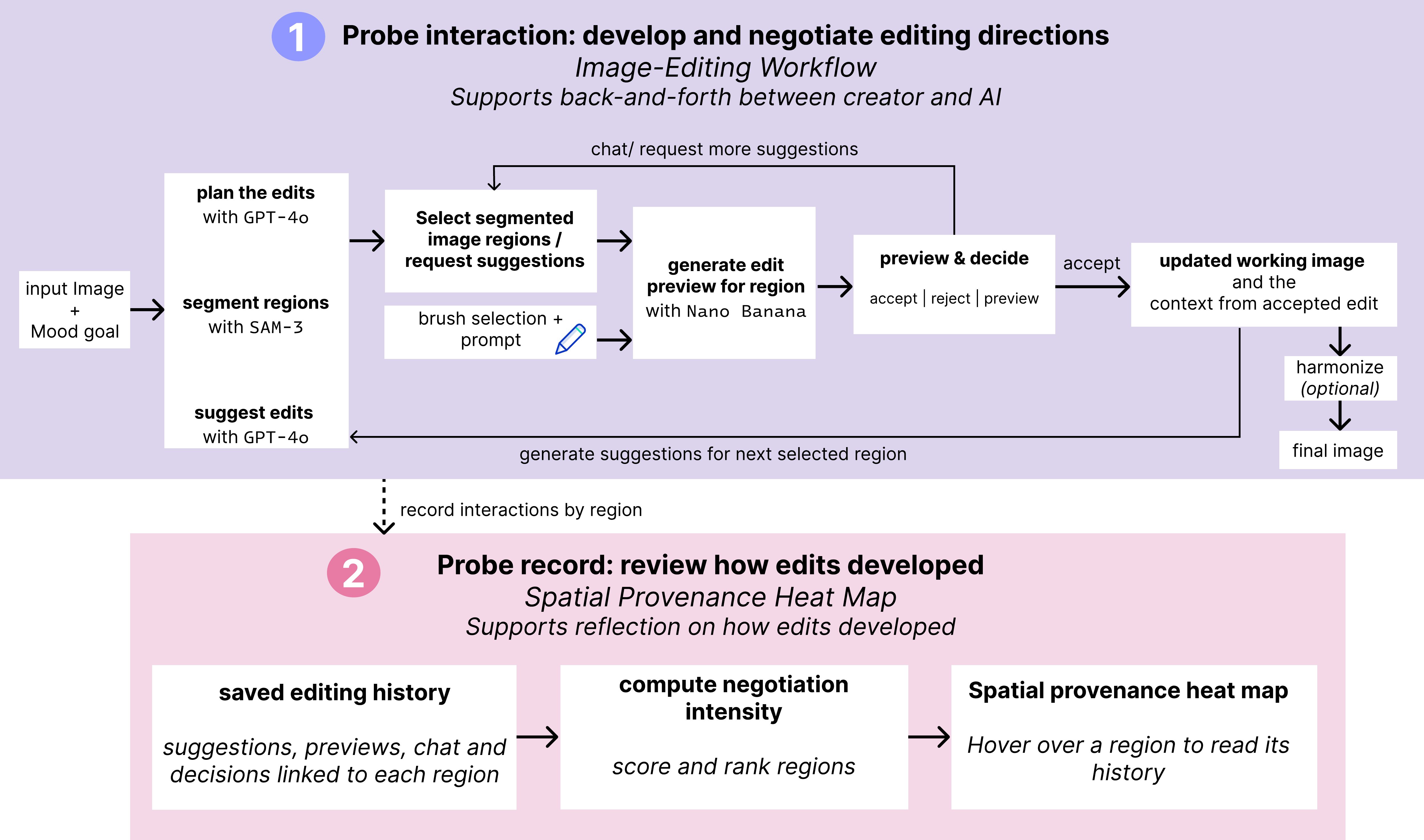}
    \caption{\systemName probe structure.
    The editing workflow supports both AI-suggested and user-initiated directions so participants can negotiate how edits develop. The resulting suggestions, revisions, previews, and decisions are recorded by region and later presented through the spatial provenance heat map for reflection. Roop is used as a design probe rather than evaluated as a technical system.}
    \Description{A two-part system diagram. The upper panel shows
    image and mood input, planning with GPT-4o, segmentation with
    SAM3, regional suggestions, preview generation with Nano Banana,
    creator review, and image updates. Feedback loops support
    requesting alternatives and generating subsequent suggestions.
    Brush edits provide an alternative input, and optional
    harmonization produces a finished image. A logging connection
    leads to the lower panel, where saved regional interactions
    are scored and ranked to produce a heat map with editing
    histories available on hover.}
    \label{fig:architecture}
\end{figure*}

\systemName updates subsequent suggestions using the developing image and previously accepted edits, as shown by the feedback loop in Figure~\ref{fig:architecture}. Accepted changes are incorporated into the working image, and descriptions of accepted edits are appended to the session context supplied to later generations. Suggestions for affected regions are regenerated when the user next opens them.

For example, accepting a brighter sky can change the context used to generate suggestions for the water. Later suggestions might then include reflections that fit the revised lighting. This allows editing to proceed in response to the developing image rather than treating each region as an independent task.

\subsubsection{User-initiated edits and harmonization}
At any point, Sophia can also initiate her own edits without first choosing an AI suggestion. The brush tool (Figure~\ref{fig:system1}~C) lets her paint over the area she wants to change and describe the edit in a text prompt. For example, she paints over the dog and enters \textit{“add a hat and sunglasses”}. The edit is generated using Nano Banana and applied to the working image. For brush-initiated edits, Sophia selects the region and supplies the editing instruction. These edits become part of the working image, allowing them to shape subsequent AI suggestions for other regions too.

Once Sophia is satisfied with her edits, she clicks Harmonize (Figure~\ref{fig:system1}~F), which generates three options for blending the individual edits into a coherent final image.  
Sophia selects the result that best matches her intent and can further refine it through chat if needed (Figure~\ref{fig:system2}~G). Harmonization can introduce global changes, like lighting and contrast adjustments, which are not captured in the spatial provenance record. We discuss this limitation in Section ~\ref{sec:lims_fut_work}.

\subsection{Spatial Interaction Provenance Heat Map}

After completing her edits, Sophia opens the provenance heat map ~(Figure ~\ref{fig:system2}~H). The heat map summarizes negotiation intensity: the amount of recorded back-and-forth around each edited region, based on actions such as chat, previewing alternatives, requesting similar suggestions, rejecting, and applying edits (details in section \ref{sub:negotiationintensity}). Cooler colors indicate regions with less human negotiation, while warmer colors indicate greater human negotiation relative to the other scored regions in the same session. Negotiation intensity is a summary of recorded engagement, not a measure of creative contribution. The system assigns brush-initiated edits maximum intensity because Sophia selected the region and supplied the editing instruction.

The heat map colors the \textit{regions} where edits were generated and negotiated, rather than only the pixels that visibly changed in the image. As a result, a colored region can include areas that remain visibly unchanged from the original. 
% The map refers to the image before harmonization, so changes introduced during harmonization are not represented.

Sophia can hover over a region to revisit its negotiation history, including \systemName's suggestions, her chat with \systemName, and the suggestions she previewed. This connects each recorded region back to the interactions through which its edit was developed. 
% \systemName summarizes these recorded interactions using a measure we call \textit{negotiation intensity}. Negotiation intensity summarizes recorded engagement within each region. The accompanying history lets Sophia inspect the interactions behind that summary.
% For each edited region, the heat map encodes the negotiation intensity  - measure of how actively the human steered each region's creative direction  - as a color gradient from blue (mostly AI-driven) to red (mostly human-driven). This lets Sophiacompare her felt sense of contribution against the behavioral record of what actually happened.
% She could have also viewed the heat map while actively editing her image and see it update in real-time, but in this instance she chose not to, to avoid being influenced by what she saw while editing.

% We discuss \textit{Negotiation Intensity} as the currency for the provenance heat map, which serves as a proxy for creative decision-making and creative contribution during the editing process. We examine how participants responded to this view in Section \ref{sections/05_findings}.

\subsubsection{\textit{Negotiation Intensity}}
\label{sub:negotiationintensity}

For each scored region $i$, \systemName computes \textit{negotiation intensity} from 5 recorded interaction signals: number of user chat turns ($T_i$), number of edit suggestions previewed on the canvas ($V_i$), number of \textit{More like this} requests ($M_i$), whether a suggestion was rejected ($R_i=1$ if rejected; $R_i=0$ otherwise), and whether a suggestion was applied  ($A_i=1$ if applied; $A_i=0$ otherwise). These signals capture the ways users can respond to and shape an AI suggestion within \systemName: engaging through chat, exploring alternatives, requesting similar options, rejecting a suggestion, or applying one. Each signal contributes equally to the composite score:

$$c_i = T_i+V_i+M_i+R_i+A_i$$

We use unweighted sum because we lack an empirical basis for assuming different weights. This provides a consistent summary of recorded actions, but does not establish that those actions have equal creative significance. \systemName then ranks regions by their composite score within each editing session and normalizes their rank:
$$t_i = \frac{rank(c_i)}{n-1},$$
where $rank(c_i)$ is the zero-based position of region $i$ when regions are sorted in ascending order by composite score, and $n$ is the total number of scored regions. When $n=1, t_i=1$. Ties are broken using elapsed time, measured from when a region was presented to when the user made a decision. Elapsed time does not enter the composite score, but can affect the final ordering when scores are equal. The colors show which regions had more or less recorded activity compared with others in the session. Even when two regions have similar scores, they can appear at opposite ends of the color scale if they are the only two scored regions. We discuss how this may affect interpretation of the map in Section \ref{sec:lims_fut_work}.

The heat map ranks regions relative to one another within the same editing session. Warmer colors indicate higher ranked engagement among scored regions, rather than a measure of creative contribution. Brush-initiated edits receive maximum intensity as a design choice because the user selects the region and supplies the editing instruction. This assignment treats user initiation as a strong signal of engagement, without accounting separately for visual details supplied by AI. In the study, participants compared their own annotations with this representation and explained where its assumptions did or did not match their understanding of contribution.

\section{Study Design}

We conducted a within-subjects study with 20 participants to examine how judgments of creative contribution differed between user-initiated and AI-suggested editing workflows (RQ1), and how they interpreted a spatial provenance record of their own editing process (RQ2). Each participant completed two editing conditions, with condition order counterbalanced across participants:

\begin{itemize}
\item \textbf{User-initiated editing:} Participants used a simplified version of the same interface where AI did not suggest edits. Instead, participants selected regions themselves, typed a description of what they wanted, and AI generated the requested changes. No AI suggestions were surfaced unprompted.
\item \textbf{\systemName (AI-suggested editing):} Participants used \systemName as described in Section~\ref{sec:probe}. AI suggested regional edits toward their mood goal, and participants could accept or reject suggestions, explore alternatives, redirect them through chat, or initiate their own brush edits.
\end{itemize}

Both conditions used the same image-generation infrastructure and two compositionally similar starting images, each showing a figure in the foreground against a distinct background. Participants produced different outputs in the two conditions, so their ratings could also reflect differences in the results they obtained. Mood goals were also drawn from a provided list rather than held constant across conditions. We used a curated list rather than fully open-ended goals so participants could quickly form a concrete, personally legible sense of what they wanted the image to convey, closer to how a creator might pursue a real mood-driven edit than an abstract prompt would allow. As a result, the specific goal, image, and condition combination differed for each participant across the two tasks, so ratings may reflect differences in the goals and edits pursued and not only the presence of AI suggestions.

After completing both tasks, participants annotated their contribution to the AI-suggested output before reviewing its spatial provenance record. This allowed us to examine their initial attributions and their explanations when comparing those attributions with the recorded history.

% \begin{table*}[!ht]
% \caption{}
% \label{tab:participants}
% \small
% \begin{tabular}{lllp{5.5cm}}
% \toprule
% PID & Experience & Yrs.\ editing & Prior AI image editing tools \\
% \midrule
% P1  & Novice & $<$1 & None \\
% P2  & Novice & $<$1 & ChatGPT \\
% P4  & Novice & $<$1 & None \\
% P7  & Novice & $\sim$1.5 & ChatGPT, Claude \\
% P10 & Novice & $<$1 & Firefly (image generation only) \\
% P11 & Novice & $<$1 & ChatGPT \\
% P16 & Novice & $<$1 & None \\
% P17 & Novice & $<$1 & None \\
% P19 & Novice & $<$1 & ChatGPT \\
% P20 & Novice & $<$1 & ChatGPT (recommendations only) \\
% P6  & Intermediate & 2--4 & None \\
% P13 & Intermediate & 2--4 & Claude, Adobe Express \\
% P14 & Intermediate & 2--4 & ChatGPT \\
% P3  & Experienced & 10+ & Firefly, Lightroom AI, ChatGPT, diffusion models \\
% P5  & Experienced & 5--6 & Firefly, LoveArt \\
% P8  & Experienced & 10+ & AI background removal tools, BeautyPlus \\
% P9  & Experienced & 5+ & Photoshop Generative Fill \\
% P12 & Experienced & 20+ & Gemini \\
% P15 & Experienced & 17  & Midjourney, ChatGPT, Gemini \\
% P18 & Experienced & 5+ & GPT-4o API, Stable Diffusion API \\
% \bottomrule
% \end{tabular}
% \end{table*}

\begin{table*}[t]
\centering
\caption{Participants' reported image-editing experience and prior use of AI tools for image-related tasks, ordered by participant ID. Researchers assigned experience categories using years of image-editing experience, including non-AI tools: novice ($<2$ years), intermediate ($2$--$<5$ years), and experienced ($\geq5$ years).}
\label{tab:participants}
\small
\begin{tabular}{@{}lllp{7cm}@{}}
\toprule
Participant & Experience category & Years editing & Prior AI tools used for image-related tasks \\
\midrule
P1  & Novice       & $<1$          & None \\
P2  & Novice       & $<1$          & ChatGPT \\
P3  & Experienced  & $10+$         & Firefly, Lightroom AI, ChatGPT, diffusion models \\
P4  & Novice       & $<1$          & None \\
P5  & Experienced  & $5$--$6$      & Firefly, LoveArt \\
P6  & Intermediate & $2$--$4$      & None \\
P7  & Novice       & $\approx 1.5$ & ChatGPT, Claude \\
P8  & Experienced  & $10+$         & AI background removal tools, BeautyPlus \\
P9  & Experienced  & $5+$          & Photoshop Generative Fill \\
P10 & Novice       & $<1$          & Firefly (image generation only) \\
P11 & Novice       & $<1$          & ChatGPT \\
P12 & Experienced  & $20+$         & Gemini \\
P13 & Intermediate & $2$--$4$      & Claude, Adobe Express \\
P14 & Intermediate & $2$--$4$      & ChatGPT \\
P15 & Experienced  & $17$          & Midjourney, ChatGPT, Gemini \\
P16 & Novice       & $<1$          & None \\
P17 & Novice       & $<1$          & None \\
P18 & Experienced  & $5+$          & GPT-4o API, Stable Diffusion API \\
P19 & Novice       & $<1$          & ChatGPT \\
P20 & Novice       & $<1$          & ChatGPT (recommendations only) \\
\bottomrule
\end{tabular}
\end{table*}

\subsection{Participants}

We recruited participants with any level of image-editing or AI-tool experience, through company-wide advertisements at a large software company. The sample included 20 participants with varying experiences (Table~\ref{tab:participants}). We assigned experience categories based on participants’ reported years of using image-editing tools, whether or not those tools involved AI: novice (fewer than two years), intermediate (at least two but fewer than five years), and experienced (five or more years). The sample included 10 novice, three intermediate, and seven experienced participants. Sessions were conducted in person and lasted approximately 60--75 minutes. Participants received a \$30 gift card.

\subsection{Procedure}

Each session included four stages. Condition order and starting image were independently counterbalanced across participants, so which image a participant saw was not confounded with which condition (user-initiated or \systemName) came first.

\subsubsection{\textbf{Introduction and background.}}
The researcher explained the study’s focus, and participants gave informed consent for screen and audio recording. Participants described their background with image editing, prior experience with AI tools, and what makes creative work feel like their own (Appendices~\ref{appendix:intro} and~\ref{appendix:warmup}).

\subsubsection{\textbf{Editing tasks and post-task reflection.}}
Participants spent approximately 15 minutes editing in each condition. They used a different starting image and selected a new mood goal from a provided list for each task. After each task, participants rated their perceived role in the creative decisions on a seven-point scale, where 1 indicated \textit{AI made all the creative decisions} and 7 indicated \textit{I made all the creative decisions}. They then answered open-ended questions about their felt creative ownership during the editing process. After completing both conditions, participants compared their experiences with the two workflows (Appendices~\ref{appendix:posttask} and~\ref{appendix:comparative}).

\subsubsection{\textbf{Contribution annotation.}}
Participants spent approximately 10 minutes annotating their \systemName output image using a digital annotation tool.  They circled regions in red for primarily their own creative decisions, blue for primarily AI-led decisions, or purple for collaboration, and explained their choices aloud. This annotation captured participants’ self-reported attribution of creative contribution \textit{before} they saw the spatial provenance heat map (Appendix~\ref{appendix:annotation}).

\subsubsection{\textbf{Provenance review and interview.}}
\label{sec:4.2.4}
Participants then spent approximately 10 minutes inspecting the heat map and its associated regional histories alongside their annotations. During review, we described the colors using language about human and AI creative effort. Participants therefore encountered the map as both a summary of recorded interaction and a representation inviting judgments about contribution. We asked what matched their recollection, what surprised them, whether they would change their annotations, and what the representation captured or missed. The session concluded with questions about their overall experience of reviewing the record (Appendix~\ref{appendix:heatmap}). This discussion elicited both agreements and disagreements, including occasions when participants remembered additional input, reconsidered an attribution, or retained their original judgment.

\subsection{Analysis}

We summarized the paired creative-decision ratings and counted how many participants rated their contribution lower, higher, or the same in the AI-suggested condition. We compared the paired ratings using the Wilcoxon signed-rank test.

We analyzed the transcripts using reflexive thematic analysis~\cite{braun_clarke_reflexive_2019}. One author coded all 20 transcripts inductively. Two additional authors each independently coded five transcripts, with no overlap between their subsets, and discussed their interpretations with the first author. These discussions were used to refine codes, identify new themes, and examine contradictory cases.

We also compared accounts of specific edits across the editing, annotation, and provenance-review stages. We examined when participants distinguished contributions within an edit, revised their judgments, or kept the same attribution even after recognizing additional recorded activity. These comparisons helped distinguish discrepancies concerning the record’s coverage or display from differences in how participants interpreted contribution. These were not mutually exclusive: a participant could question the scoring rule while also explaining why their recorded actions did not match how they understood their contribution. We interpreted these accounts in relation to both the probe’s scoring rules and the contribution-related language used during review.

\section{Findings}
% how do we see AI stuff differently? read about history of automation - ground framing in some theory - the process of creation > the outcome -- 
% can we "sort of" direct an outcome? 
% humans look at stuff and judge the output, generally
% as researchers, we know that there's a process behind the output, which is important
% how we evaluate the end outcome depends on how the output was developed
% if ai is just a medium - then how do you capture contribution when the whole thing is ai generated
% ansel prepping everything for years before capturing it - art is more than just what you see on the canvas 
% something about "i directed the outcome" is an interesting framing

We first examine how participants understood their own and AI’s creative contributions, including how their judgments differed between user-initiated and AI-suggested editing (RQ1). We then examine how participants interpreted the spatial provenance record in relation to those judgments (RQ2). Our analysis distinguishes the decisions participants attributed, the meaning they gave their interactions, and what they reconsidered after reviewing the history.
\begin{figure}[!ht]
  \centering
  \includegraphics[width=\textwidth]{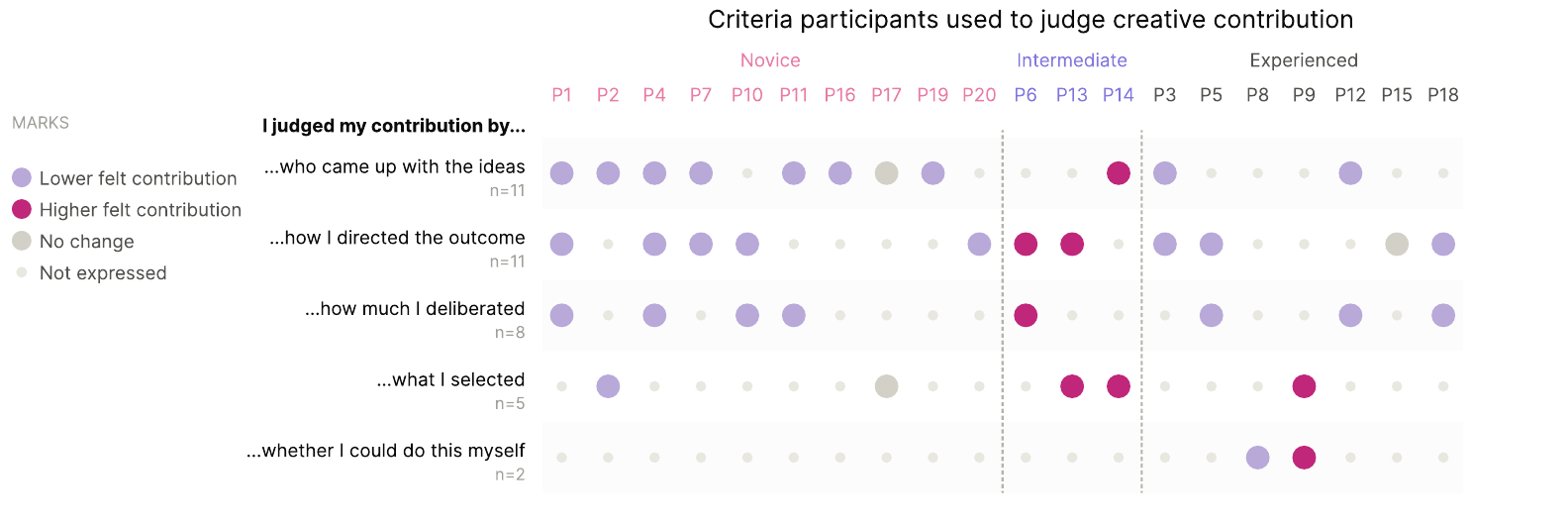}
  \caption{Criteria participants used when judging their own and AI’s creative contributions (N=20). Each column represents one participant, and a filled circle indicates that the participant expressed that criterion during the study. Participants could use more than one criterion. Circle color shows whether the participant reported lower, higher, or unchanged self-attributed contribution in the AI-suggested condition compared with user-initiated editing.}
  \label{fig:frameworks}
\end{figure}
\subsection{Creators judged contribution through the decisions they believed they shaped}
\begin{figure}[!ht]
  \centering
  \includegraphics[width=\textwidth]{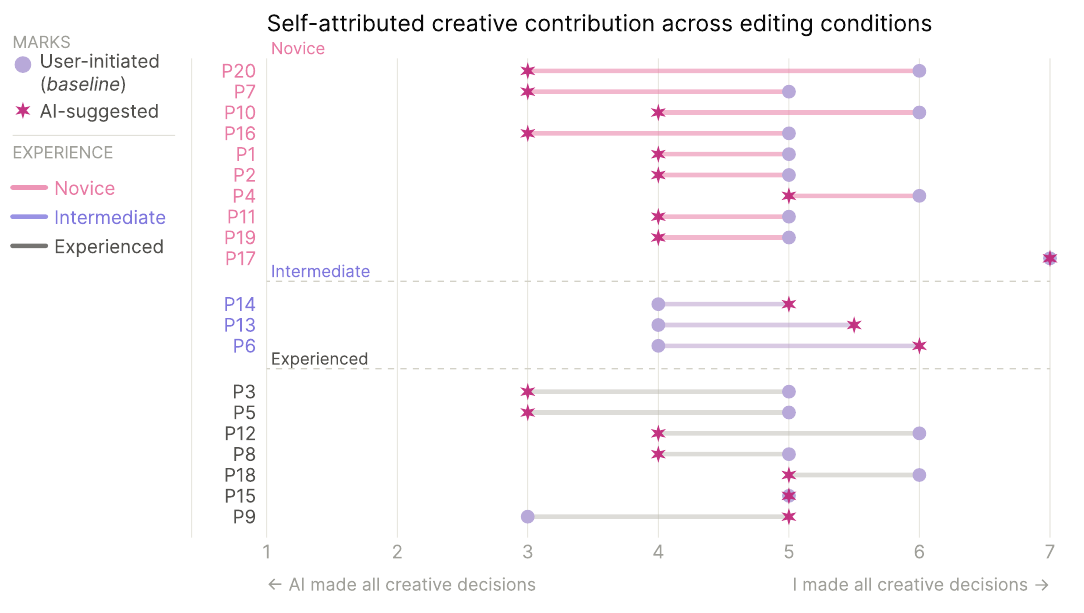}
  \caption{Self-attributed creative contribution for all 20 participants in user-initiated and AI-suggested editing. Each row shows one participant’s ratings in both conditions, connected by a line. Fourteen of 20 participants rated their own creative contribution lower in the AI-suggested condition, four rated it higher, and two reported no change. Participants are grouped by prior image-editing experience.}
  \label{fig:ownership}
\end{figure}

As seen in Figure \ref{fig:frameworks}, participants described contribution in terms of originating ideas (P1, P3, P4, P19), directing changes (P5, P10, P15, P16), and selecting or refining what AI produced (P6, P9, P13, P18). Participants often drew on more than one of these considerations. Participants also distinguished their contribution to an overall image from their contribution to a particular detail (P6, P8, P13, P15, P17) such as an object's texture, color, or pattern.
Self-attributed creative contribution was generally lower in the \systemName condition. Fourteen of 20 participants rated their contribution lower when AI suggested edits, four rated it higher, and two reported no change (Figure \ref{fig:ownership}). 
% Fourteen of 20 participants rated their contribution lower when AI suggested edits, four rated it higher (P6, P9, P13, P14), and two reported no change (P15, P17; Figure \ref{fig:ownership}).
Median ratings were 5 in user-initiated editing and 4 in AI-suggested editing (Wilcoxon signed-rank test, W=40.5 and p=.046; paired rank-biserial correlation, r=.53, with positive values indicating higher user-initiated ratings). These ratings compare the two editing workflows, which also differed in the images, selected goals, and resulting edits.

Participants who preferred initiating edits often emphasized deciding what to change before seeing alternatives. P4 felt more creative when “I had to decide which elements to change,” while P3 described their role with AI suggestions as “my job is to approve or disapprove.” For P3, approving a direction involved less personal thinking than deciding how to approach the image themselves.

Receiving suggestions did not, however, prevent participants from identifying their own contribution. P6 valued selecting among proposals and “having the last bit of revision.”
% P9 felt more able to “specify my intent more clearly” in the suggested-editing workflow, whereas the user-initiated workflow had made it difficult to refine details and preserve other parts of the image. 
% P9’s account concerned the control available in the workflow, rather than attributing this benefit specifically to receiving AI suggestions.  \textcolor{magenta}{CD: is this because of the AI-suggestions? or because of the built-in image segmentation? might be worth clarifying}
P13 similarly valued choosing and adjusting the scene while letting AI supply smaller details. These accounts tied contribution not only to who introduced a suggestion, but also to what participants did with it.

Participants’ intentions could also develop through those suggestions. P19 already wanted to change the person's outfit in the input image, but seeing alternatives helped them realize that they wanted to change both the top and pants. P10 described choosing yellow boots after seeing AI generate a yellow dress, explaining that the dress inspired the color choice while still considering the decision to add yellow boots their own.

Participants also distinguished contribution from preference for the result. P4 considered the AI-suggested image better even though they felt they had contributed more to the other image. P20 likewise preferred the more AI-driven result while reporting greater control in user-initiated editing. Lower self-attributed contribution therefore did not necessarily indicate a less preferred or less enjoyable outcome.

\subsection{A single edit could involve contributions at different levels}
\label{sec:5.2}

\begin{figure}[!ht]
  \centering
  \includegraphics[width=0.9\textwidth]{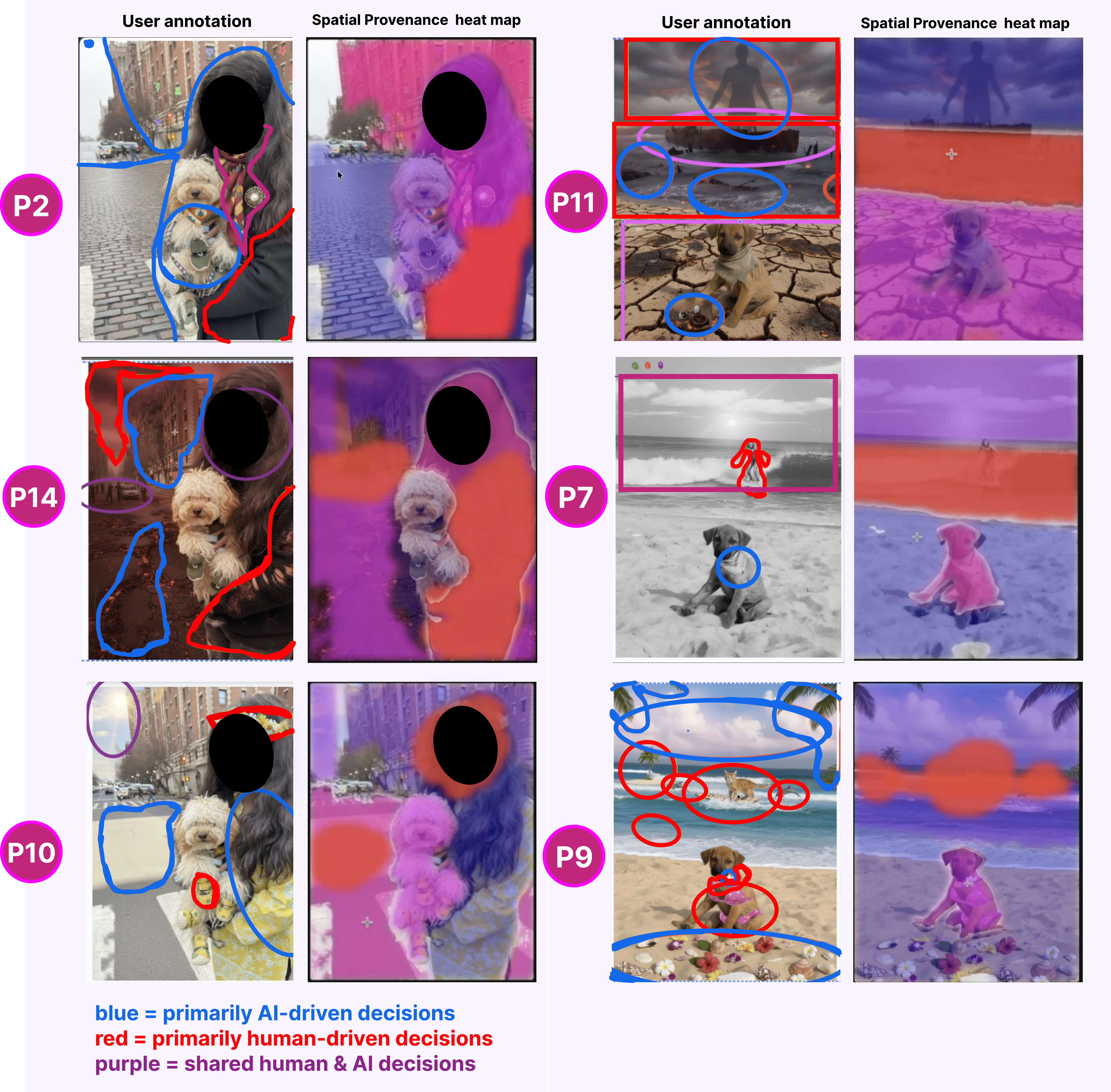}
  \caption{Participants’ self-annotated contributions (left in each pair) and \systemName’s spatial provenance heat maps (right). Annotation colors are standardized: blue indicates primarily AI-driven decisions, red primarily human-driven decisions, and purple shared decisions. Heat-map colors show relative recorded negotiation from blue (lower) to red (higher), with brush-initiated edits assigned maximum intensity. Reviewing the history led P2 to reconsider the buildings as more collaborative, while P10 recalled input to the road but retained AI attribution. Faces are obscured to preserve anonymity. 
  % \textcolor{magenta}{CD: I find it a little confusing that red in the left image means "more AI" and red in the right image means "more human". can we standardize?}
  }
  \label{fig:heatmap-comparison}
\end{figure}

Eleven participants distinguished their role in specifying an edit from AI’s role in determining how it appeared (P5, P6, P8, P9, P11, P13, P14, P15, P17, P19, P20). These distinctions concerned different decisions within the same object or edited region. P9 distinguished choosing a lei to put on the person in the foreground from AI choosing its specific flowers (Figure \ref{fig:heatmap-comparison} P9), while P14 explained, “I wanted zombies. I didn't know what specific kind of zombies (Figure \ref{fig:heatmap-comparison}~P14).” P17 similarly wanted sandals for the dog without having a particular appearance in mind. Participants could therefore attribute the broader change to themselves while acknowledging AI’s contribution to its visual realization.

These distinctions informed how participants attributed the resulting edit. P11 considered a generated ship collaborative because they had introduced the idea, while AI supplied possibilities for its form and wreckage that they had not considered: “I think AI definitely made some design choices (Figure \ref{fig:heatmap-comparison} P11).” P6 similarly distinguished deciding to add water from determining its style, but emphasized that the appearance did not match what they had wanted. They explained, “I feel like AI took that part of this design decision,” and consequently felt that the result belonged less to their own work. Acknowledging a human-originated idea therefore did not necessarily lead participants to attribute the resulting edit primarily to themselves.

Comparing annotations with the provenance map showed how these distinctions affected interpretations of the same recorded activity. P8 had requested changes to a person but regarded the result as collaborative because AI determined details such as the clothing pattern, hair length, and accessory placement. When they saw the strongly human-marked region in the spatial provenance heat map, they recognized the connection to their brush selection but questioned what that assignment represented: “the heat map does capture my actions very well,” but it did not capture the smaller design choices they attributed to AI. The location of their activity corresponded to the record, while their attribution also considered other decisions made within that location.

P13’s account followed a similar pattern. They had specified wanting to add cement and wreckage to the background for an apocalyptic vibe, but expected more shared contribution than the map showed. During reflection, they recognized the heat map as “very spot on of what I touched.” However, they also called its calculation “too simple of a calculation.” AI had determined how the cement looked and what it contained: “that's not something I would exactly picture.”  When asked whether they would change their annotation (how they interpreted their contribution), P13 declined. The heat map captured where they had acted, but this did not settle how they divided contribution between their instructions and the details AI supplied.

Together, these accounts show that attributing an edit involved both distinguishing who shaped its constituent decisions and considering what those decisions meant for the result. Participants could acknowledge their own role in initiating or specifying a change while treating AI’s contribution to its appearance as grounds for shared or more AI-driven attribution. The maximum intensity assigned to brush edits made this distinction particularly visible. P8 and P13 questioned that assignment while recognizing the actions it represented. Their accounts show what this scoring rule left out: requesting a change did not mean they had determined all the details of its appearance.

\subsection{More interaction did not necessarily mean greater creative contribution}
\label{sec:5.3}

Seven participants explained their contribution by considering what their interactions expressed or accomplished, including specifying an intention, developing a direction, or attempting to correct an unwanted result (P5, P6, P10, P14, P16, P19, P20). These accounts gave interaction different meanings: effort could support a claim of creative contribution, but additional activity did not necessarily make the resulting edit feel more human-driven.

For P16 and P20, working through difficulties supported their sense of contribution. P16 felt like “I was doing the heavy lifting” when they described repeatedly having to redirect unsuitable suggestions for editing the street. P20 similarly felt that edits requiring repeated correction or refinement reflected more of their own contribution because they had to think through how to get the result they wanted. In these accounts, effort mattered as part of expressing and pursuing what participants wanted.

P10 described a different relationship to repeated correction. Although the road involved considerable back-and-forth, P10 still considered the final road primarily AI-driven after becoming frustrated and accepting the result: “you win, like you can have your say in like what this road looks like (Figure \ref{fig:heatmap-comparison}~P10)..” P6 likewise explained that “the effort I put in did not converge.” They had sought a more specific appearance than AI delivered and therefore felt less personal contribution to the result.

The distinction also extended beyond correction. P19 recognized spending time exploring building alternatives but continued to attribute the edit primarily to AI because they had not introduced its idea. Conversely, P5 felt that their earlier instruction to add snow was underrepresented despite accepting the result with little further negotiation. P14 distinguished pursuing their own intention from converging with AI on a direction. P16 and P10 described similar amounts of effort, but they interpreted that effort differently. P16 saw the repeated revisions as continuing to shape the edit, while P10 saw them as failed attempts before eventually accepting AI’s result.

Together, these accounts show why interaction amount did not fully explain attribution. Participants considered whether their actions expressed an existing intention, helped them develop a direction, or ended in accepting a result they had not wanted. Their explanations concerned the relationship between their actions and the result, not only the amount of activity recorded. P16 treated continued correction as evidence of directing the edit, while P10 described accepting the road after giving up on further correction. These contrasting accounts explain why repeated interaction could support different judgments of contribution.

\subsection{Reviewing the history helped participants remember their input, but did not always change how they judged their contribution}

Six participants reconsidered input they had overlooked or where an idea had come from after reviewing the history (P2, P5, P6, P7, P10, P11). These reconsiderations took different forms.

For some, recalling forgotten participation led them to revise their contribution annotation. P2 had annotated the buildings as mostly AI-driven, but reviewing the history reminded them of specific opinions they had expressed about how the buildings should look. As they explained, "I guess I just didn't remember it well enough," and said they would make the region "more purple" to reflect the collaboration. P6 had a similar reaction to a region: seeing the record, they said, "I totally forgot I had added this part for the car... I would have circled it if I remembered."

For others, recalling forgotten participation did not change their attribution. After seeing the spatial provenance heat map, P10 remembered that they had previously requested a detail they had forgotten by the time they annotated the image. However, they still chose not to revise their contribution annotation. As described in Section \ref{sec:5.3}, P10 had become frustrated with the road edit and eventually stopped trying to direct it. Remembering an additional request therefore did not change their view that the final road was primarily shaped by AI.

Not every disagreement involved forgetting. As described in Section \ref{sec:5.2}, P13 recognized where they had acted but questioned what those actions meant for contribution. Others found the heat map useful without changing their account: P20 reported no change or surprise, while P1 said the record “helped me better to communicate my creative process.”

Participants could also agree with the map overall while disagreeing with how it represented particular edits. P14 described their annotations as being in "maybe some vague 75\%, 80\% agreement'' with the heat map, but questioned its treatment of both the sky and zombies. For the sky, they explained, "I had a vision, but then it just says it's like leaning towards AI.'' They raised the opposite concern about the zombies: "it gave me the credit for zombies,'' although they had not had a specific appearance in mind. Their overall agreement did not mean they accepted how the heat map represented contribution to every edit.

These accounts show that the history could help participants remember and explain their involvement without requiring them to agree with the map. P2 changed how they attributed the edits to the buildings after remembering additional input, while P10 remembered an earlier request but still considered the edited road as primarily AI-driven. These responses occurred during the combined annotation, provenance-review, and interview process; they do not isolate the effect of the heat map alone.

\section{Discussion}

Our study examined how creators attributed contributions to images they had edited and how those judgments related to a spatial record of their interactions. The findings reveal two sources of disagreement with the display. Some concerned how the probe recorded or summarized activity. Others concerned how participants interpreted actions they recognized as their own. These sources could overlap, but they have different implications: changing the display may address a misleading summary, while understanding attribution also requires considering what participants believed their actions contributed.

\subsection{Locating an edit does not fully explain its contribution}

Prior work has shown that attribution can depend on contribution type, initiative, and the distinction between human ideas and AI-supplied details \cite{he2025contributionsdeservecreditperceptions,gjing2026man,huang_copyrighting_2025}. Our study shows that these contribution distinctions do not always align with the units captured by provenance records. By comparing creators’ own annotations with the histories of edits they made themselves, we found that a single edited region could contain several differently attributed decisions.

Connecting interaction history to edited image regions made these comparisons concrete. Participants could identify the part of the image they were discussing and inspect the interactions associated with it. However, a region could contain several differently attributed decisions. P8 and P13 recognized that they had decided what to change, while AI determined some of the details of how that change appeared. Both contributions could therefore occur within the same edited region. Making the provenance map more spatially precise would not necessarily separate the human and AI contributions, because the difference was between creative decisions, not locations in the image.

\systemName’s treatment of brush edits also shaped this result. Assigning them maximum intensity emphasized user initiation and did not separately represent details supplied by AI. Participants’ objections therefore do not establish that every spatial provenance representation would produce the same difficulty. They do, however, explain why identifying a user-initiated action was insufficient for the judgments these participants made.

An accurate provenance record can show that a creator requested an edit and received a generated result. That information does not necessarily establish which details about the edit the creator had specified or how much those details mattered to their attribution. Participants could acknowledge their own request while giving AI credit for its appearance. Their explanations identify considerations that the regional activity summary did not express.

The interaction history shows what happened. It does not always explain what the creator thinks they contributed. That may require the creator to explain which decisions they see as their own and which they attribute to AI.

\subsection{Directing creative work can involve developing an intention}

Participants’ accounts suggest that creative direction does not need to begin with a complete specification of the result. Schön describes design as a reflective conversation with materials, in which people develop their understanding by making moves and responding to their consequences \cite{schon_reflective_1983}. P19’s developing clothing choices and P10’s choice of yellow boots illustrate this process in AI-assisted editing: participants acknowledged that suggestions influenced their thinking while identifying subsequent decisions as their own.

These accounts help explain why the immediate initiator of an edit does not settle its attribution. A human-initiated request may develop from an earlier AI suggestion, while revising an AI proposal may help a creator determine what they want. Recording that sequence preserves useful evidence, but participants’ explanations establish how they understood the relationship between those events and their own direction.

% Participants’ accounts suggest that creative direction does not need to begin with a complete specification of the result. Schön describes design as a reflective conversation with materials, in which people develop their understanding by making moves and responding to their consequences \cite{schon_reflective_1983}. This provides a useful perspective on participants who became more specific about what they wanted after seeing AI-generated alternatives.

% For example, P19’s existing intention to change an outfit developed as they considered how suggested clothing worked with the rest of the image. P10 described choosing yellow boots in response to an AI-generated yellow dress. Such accounts complicate a simple distinction between expressing a fully formed human idea and adopting an AI idea. A creator can acknowledge AI’s influence while identifying subsequent decisions as their own.

% This perspective also helps interpret the effort findings. Participants considered whether their interventions expressed or developed a direction, rather than treating all intervention as equivalent. P16’s continued correction supported a claim of direction, while P10’s acceptance after frustration did not carry the same meaning. Understanding contribution therefore requires considering how participants relate their activity to the developing result, alongside who initiated the interaction.

\subsection{Provenance can support reflection without settling attribution}

HaLLMark demonstrated how artifact-linked provenance could support writers’ awareness of AI use and reflection on their work \cite{hoque2024hallmark}. In our study, participants first marked which parts of the image they saw as human-, AI-, or jointly shaped, and only then reviewed the provenance record. We then asked them to compare the two. This made it possible to distinguish remembering additional participation from changing how they attributed contribution.

P2 reconsidered the edits they made to the building as more collaborative after the heat map reminded them of specific choices they had made about how the buildings should look. P10 also remembered a specific request they had made for the road, but still considered the final road mainly AI-driven because they had ultimately accepted a result they did not want. Remembering more of the process did not necessarily change how participants judged contribution.

These cases show why recall and attribution should be examined separately when evaluating provenance interfaces. We do not claim that the visualization alone caused participants to change their judgments, since the annotation task and interview questions also prompted reflection.

Agreement with the heat map did not always mean agreement with what it implied about contribution. P13 recognized where the map recorded their activity while questioning the contribution it suggested, and P14 broadly agreed with the map while disputing specific regional assignments. Our study shows why these aspects should be separated: recognizing where someone acted, how much they interacted, and which decisions they considered their own were not always the same judgment.

A provenance interface may therefore still be useful even when it does not change someone’s attribution. It can help creators remember what they did or explain why they see an edit as human-, AI-, or jointly shaped. These outcomes should be considered separately from whether participants agree with the display’s implied allocation of contribution.

\subsection{Preserve context to help creators reconstruct and communicate contribution}

Our findings suggest that provenance systems should preserve not only what happened, but enough context for creators to interpret how different interactions shaped the work. In participants’ accounts, this context included what they were trying to achieve, which decisions they considered their own, and whether their interventions helped develop a direction or ended in accepting an unwanted result.

This leads to two design implications. First, systems should be clear about what the visualization is actually showing. If a heat map summarizes things like chat, revisions, previews, or brush edits, those should be presented as recorded activity rather than as a direct measure of creative contribution. The display should also make its scoring rules visible, such as assigning maximum intensity to brush-initiated edits. This would help creators distinguish what the system recorded from what they believe they contributed. We did not test a different presentation, so whether clearer labeling would change these judgments remains an open question.

% This leads to two design implications. First, systems should represent recorded engagement as engagement. \hl{Participants' own accounts sometimes anticipated this concern. P18 distinguished two kinds of effort behind their editing: "the mental effort" — deciding on a direction, recognizing when a suggestion fit what they wanted — and "the physical effort" — the time spent prompting, retrying, and iterating. They argued the record captured the latter well but not the former: "it indicates like this to be like my... high[ly] human, but actually I don't think it's more human. It involves like not so much like the mental effort." This distinction is close to the one we draw between recorded engagement and creative significance (Section 6.1), suggesting that creators themselves may already hold a working model of this gap even before seeing a record built on activity counts}

% The interface should distinguish what an interaction establishes from what remains open to interpretation. For example, a brush instruction establishes that the creator requested a regional change; it does not establish that they determined every detail of the result. A summary of prompts, previews, or revisions can help creators locate activity worth revisiting, but should not present that activity as a measured share of authorship or credit. Its labels and explanations should make clear what was recorded and how the summary was calculated.

Second, systems could connect selected interactions to creators’ explanations of particular decisions. For example, a creator could distinguish requesting an object from determining its appearance, or explain why repeated revisions did not make the result feel more like their own. The history would help them reconstruct the process, while the explanation would communicate what they believed their participation contributed. Adding this context during review could focus annotation on decisions creators consider important or difficult to explain, rather than requiring a justification after every action.

Future systems could also let creators explore alternative summaries of the same record. For example, they could place greater emphasis on initiating edits, selecting alternatives, or revising suggestions and inspect how those choices change the map. This could make the assumptions behind each summary visible and provide another way to explain their judgments. Changing the weights could alter how recorded interactions are summarized, but it would not capture information the system never recorded, such as an unexpressed intention or why a creator accepted a result. Customizable summaries could therefore accompany creators’ explanations, with each heat map presented as one interpretation of the recorded process.

The provenance interface should keep recorded events and creators' interpretations distinct. A creator’s explanation should remain identifiable as their interpretation, supported by links to relevant interactions. For disclosure, this suggests connecting a concise statement of AI use with relevant history and a creator’s explanation of how they shaped the work. Our study identifies this opportunity from creators’ accounts; whether such explanations improve audience understanding remains a question for a study with viewers.
\label{sec:lims_fut_work}
\section{Limitations \& Future Work}

Our study has several limitations that suggest directions for future work. First, participants completed short image-editing tasks using provided images and mood goals selected from a list. Their judgments of contribution and ownership may differ when working on their own images, pursuing their own goals, or developing work over longer periods. Participants were also recruited from one software company, which limits the range of professional and organizational contexts represented in the sample. Future work could study provenance in longer-term creative projects, including how creators interpret their histories after time has passed.

\systemName records interactions for each edited region, but does not separately represent the different contributions participants described within an edit. It also does not capture global changes introduced through \textit{Harmonize}, such as changes to lighting or contrast. The heat map refers to the image before harmonization, so its regions may not match the final output. Some discrepancies therefore came from missing or misaligned information. Future work could keep provenance records synchronized with later changes to the image, including global edits introduced through Harmonize. It could also explore ways to represent multiple human and AI contributions within the same edited region.

Negotiation intensity summarizes recorded engagement rather than measuring creative contribution or ownership. We weighted interactions equally because we lacked an empirical basis for assigning different weights, but participants did not necessarily consider these actions equally meaningful. Brush-initiated edits also received maximum intensity, even when the creator chose what to change but AI determined how the result looked. Ranking can also make differences between regions appear larger than the underlying differences in activity. Although the tasks provided time for participants to make multiple edits, this does not eliminate the effect of normalization. Some participants described regions as more or less strongly colored than expected, but their comments do not establish whether those reactions came from the ranking, the scoring rules, or their interpretation of contribution. Future work could examine whether showing different types of interaction separately, or preserving differences in activity rather than rank alone, helps creators interpret their histories.

We introduced the heat map using language about human and AI creative effort (Section \ref{sec:4.2.4}). Participants’ responses therefore reflect both the scoring rules and that framing. Some objections may have been specific to the contribution the display appeared to assign, particularly for brush-initiated edits. Although participants also explained why accurately recorded actions did not settle their attribution, we did not test whether different scoring rules or whether presenting the heat map purely as a summary of recorded interaction activity would address these concerns.

% The way we introduced the heat map may also have shaped participants’ responses. During review, we described colors using language about human and AI creative effort. This may have encouraged participants to judge whether the map gave them appropriate credit, alongside whether it captured their activity. Future work could examine how describing the colors only in terms of recorded interaction changes these judgments.

Our findings are based on how participants described their own contribution. We could not verify whether the intentions they described during the interview were already formed before they expressed them in the interface.

Participants annotated their images before reviewing the provenance record, and the interview itself also prompted reflection. Any changes in attribution therefore cannot be attributed to the heat map alone.
We also did not treat either the initial annotation or the heat map as ground truth. Instead, we used differences between them to understand how participants interpreted their own contribution.

Finally, participants saw provenance only after editing. Showing it during creation could change how they interact with AI. We also studied creators’ interpretations rather than viewers’ responses. Future work could examine what creators want to disclose and whether sharing relevant history alongside their explanations helps others understand human and AI contributions.

% * not capturing granularity - "filling in the gaps" \\
% * the equation we currently use for creating the provenance heat map is limited \\
% * harmonize feature can add hallucinated items that are not captured in the heat map. also we don't have a way to capture global edits like lighting or filter affects in the heat map.\\
% * ownership is a "felt" thing, which is intangible and cannot be directly translated into a universal metric. this suggests we might need customization for ownership representation.\\
% * viewing provenance itself can change peoples' creative behavior
% We note that brush selections do not capture execution details determined by AI during generation, like specific patterns, textures, or stylistic choices. For example, a user might use the brush to add another person in the background without specifying the race or outfit choices; these gaps are filled in by the AI. We address this limitation in Section \ref{sections/08_limitations_future_work}. Auto-accepted regions are rendered grey and excluded from the ranking, to avoid conflating absence of human engagement with low human engagement.

% * We weight these signals equally because we did not have an empirical basis for assigning greater creative significance to one form of engagement than another.

\section{Conclusion}

Participants did not always attribute contribution at the level captured by the provenance record. A single edited region could contain decisions they attributed differently to themselves and AI, and repeated revision could mean either continued direction or eventually giving up. Reviewing the history sometimes helped participants remember choices they had forgotten, but this did not always change how they judged contribution. These differences came partly from how \systemName summarized activity and partly from how participants interpreted what their actions had contributed. Future provenance tools could therefore preserve the editing history while also giving creators ways to explain which decisions they see as their own, which they attribute to AI, and why.

\bibliographystyle{ACM-Reference-Format}
\bibliography{sample-base}

\appendix

% \section{Negotiation Intensity Scoring}
% \label{appendix:scoring}

% The composite engagement score for each region is computed as:

% \texttt{score = (user turns in conversation, excluding ``show me everything'')}
% \texttt{+ variantsPreviewed.length}
% \texttt{+ moreLikeThisOn.length}
% \texttt{+ (1 if outcome === ``rejected'')}
% \texttt{+ (1 if appliedVariant is set)}

% Scores are rank-normalized within the session:
% t\_i = rank(c\_i) / (n - 1)
% where rank(c\_i) is the zero-based position of region i sorted
% ascending by score. For n=1, t\_i=1. Ties are broken by
% deliberation time (decidedAt - proposedAt).

\section{Study Protocol}
\label{appendix:userstudy}

\subsection{Introduction Script}
\label{appendix:intro}

``Thank you for joining me today. I'm studying how people think about creative 
ownership when they edit images with AI --- specifically, whose 
ideas shaped the final output and how much of the output feels 
like yours versus AI's.

Today you'll use two different image editing tools and I'll ask 
you some questions about your experience. You don't need to have 
any professional background in image editing.

Before we start, I want to ask for your permission to record this 
computer screen and audio throughout the session. The recording 
will only be used for research purposes and any personally 
identifiable information will be anonymized.''

[Wait for permission.]

``I'd also like you to think out loud as much as possible while 
you use both tools --- narrating what you're noticing, what 
you're thinking, what you're deciding, and how you're feeling, 
even if it seems obvious or not important. The more you share, 
the more helpful it will be for my research. Finally, you can opt 
out of this study at any time and skip any questions you're not 
comfortable answering. Do you have any questions before we 
start?''

%-------------------------------------------------------------------
\subsection{Warm-Up Questions}
\label{appendix:warmup}
%-------------------------------------------------------------------

\begin{enumerate}
    \item Can you tell me about your background in image editing? 
    Do you edit images regularly --- for work, personal projects, 
    or just for fun?
    \item Have you used any AI tools for image editing before? 
    If so, which ones and what was it like?
    \item When you think about a piece of creative work you've 
    made --- like a photo you've edited or an image you've created 
    --- what makes it really feel like yours?
\end{enumerate}

%-------------------------------------------------------------------
\subsection{Post-Task Questions (Repeated After Each Condition)}
\label{appendix:posttask}
%-------------------------------------------------------------------

\begin{enumerate}
    \item Looking at the final image, which parts feel most like 
    your creative contribution? Walk me through what you're 
    pointing to and why.
    \item On a scale from 1 to 7, where 1 means AI made all the 
    creative decisions and 7 means you made all the creative 
    decisions, where would you put yourself for this image? Why?
    \item Were there any moments where you felt genuinely 
    creative? What were they?
    \item Were there any moments where you felt like you were 
    just going along with what the AI gave you?
\end{enumerate}

%-------------------------------------------------------------------
\subsection{Comparative Questions}
\label{appendix:comparative}
%-------------------------------------------------------------------

\begin{enumerate}
    \item Thinking about both tools, which one did you feel more 
    like the author of the final image? Why?
    \item Did the way the AI was involved feel different between 
    the two tools? Can you describe how?
    \item In \systemName, the AI suggested changes before you had decided 
    or expressed what you wanted. Did that change how you thought 
    about the image or the creative direction? In what way?
    \item If you were describing either of these images to a 
    friend, would you describe it as your work, AI's work, or 
    something you made together with AI? Does that answer differ 
    between the two tools?
\end{enumerate}

%-------------------------------------------------------------------
\subsection{Self-Annotation Instructions}
\label{appendix:annotation}
%-------------------------------------------------------------------

``I'd like you to look at your final image from \systemName and annotate 
it. For different regions of the image, circle in \textbf{blue} 
what you feel was mostly AI, circle in \textbf{red} what you 
feel reflects mostly your creative decisions, and circle in 
\textbf{purple} what you believe was genuinely a mixed 
collaboration between you and the AI. Please explain out loud 
what you're thinking as you annotate.

Looking at your annotations overall --- does this feel like an 
accurate picture of how the image was made? Is there anything 
that surprised you about how you've annotated it?''

%-------------------------------------------------------------------
\subsection{Heat Map Reflection Questions}
\label{appendix:heatmap}
%-------------------------------------------------------------------

Participants were shown the negotiation intensity view of the 
provenance heat map, showing how much back-and-forth 
occurred between the participant and AI in each region, including 
conversational turns, alternate suggestions previewed, and 
suggestions rejected or accepted.

\begin{enumerate}
    \item What's your first reaction to this?
    \item Does it match what you remember about where you spent 
    your creative energy?
    \item Are there any regions that look warmer or cooler than 
    you expected?
    \item It says your decisions in [region X] actually shaped AI's suggestions for [region Y]. 
    Were you aware this was happening? Does knowing that change 
    how you think about your contribution to [region Y]?
\end{enumerate}

\textbf{Comparison with self-annotation:}
\begin{enumerate}
    \item Are there any regions where your annotations don't line 
    up with the heat map?
    \item If you were to re-annotate the image now that you've 
    seen the heat map, would you change anything?
    \item Does this heat map feel like an accurate account of how 
    this image was made? What does it capture well, and what does 
    it miss?
    \item Is there anything about this whole experience --- the 
    editing, comparing your annotations, or seeing the provenance 
    record --- that surprised you or made you think differently 
    about AI and creative ownership?
\end{enumerate}

\end{document}